\documentclass[twocolumn, showpacs, amsmath, amssymb]{revtex4}

\usepackage[dvips]{graphicx}

\newcommand{\average}{\langle x \rangle}
\newcommand{\cf}{{\it cf.} }
\newcommand{\diff}{{\mathrm d}}
\newcommand{\deltareli}{\delta_{\mathrm{rel, }i}}
\newcommand{\Eq}[1]{Eq.~(\ref{#1})}
\newcommand{\eg}{{\it e.g.}, }
\newcommand{\eq}[1]{eq.~(\ref{#1})}
\newcommand{\eqs}[2]{eqs.~(\ref{#1}) and (\ref{#2})}
\newcommand{\fB}{f_{\mathrm B}}
\newcommand{\fig}[1]{figure~\ref{#1}}
\newcommand{\fnstar}{f^{n*}}
\newcommand{\fN}{f_{\mathrm N}}
\newcommand{\fP}{f_{\mathrm P}}
\newcommand{\Fig}[1]{Figure~\ref{#1}}
\newcommand{\gnstar}{g_{n}}
\newcommand{\hnstar}{h_{n}}
\newcommand{\ie}{{\it i.e.}, }
\newcommand{\LN}{\mathrm{LN}}
\newcommand{\Mnt}{M_n^{\mathrm t}}
\newcommand{\N}{\mathrm{N}}
\newcommand{\oper}[2]{\mathrel{\mathop{\kern 0pt#1}\limits_{#2}}}
\newcommand{\Section}[1]{Section~\ref{#1}}
\newcommand{\Snt}{S_n^{\mathrm t}}
\newcommand{\Vnt}[1]{#1_n^{\mathrm t}}
\newcommand{\vs}{{\it vs} }
\newcommand{\xm}{{x^\mathrm{m}}}
\newcommand{\xt}{x^\mathrm{t}}
\newcommand{\Yni}{Y_{n, i}^{\mathrm t}}
\newcommand{\YnI}{Y_{n, \mathrm{I}}^{\mathrm t}}
\newcommand{\YnII}{Y_{n, \mathrm{II}}^{\mathrm t}}
\newcommand{\YnIII}{Y_{n, \mathrm{III}}^{\mathrm t}}
\newcommand{\Ynt}{Y_n^{\mathrm t}}
\newcommand{\appen}[1]{Appendix~\ref{#1}}

\newcommand{\var}[1]{{\mathrm {var}} \left(#1\right)}
\newcommand{\defin}{\equiv}

\newcommand{\Numerical}[3]{#1_{#2, \mathrm {ex}}^{\mathrm {#3}}}

\newcommand{\Proba}[2]{{\mathrm {Pr}} \left( {#1} < {#2} \right)}
\newcommand{\Repart}{{\mathrm {F}} \left( x \right)}

\newcommand{\New}[1]{{#1}}

\begin{document}

\title{Broad distribution effects in sums of lognormal random variables}

\author{M. Romeo}%
\affiliation{Institut de Physique et Chimie des Mat\'eriaux de
Strasbourg, CNRS (UMR 7504) and Universit\'e Louis Pasteur, 23 rue
du Loess, \New{BP 43, 67034 Strasbourg Cedex 2}, France.}
\altaffiliation{Corresponding author: romeo@ipcms.u-strasbg.fr}

\author{V. Da Costa}%
\affiliation{Institut de Physique et Chimie des Mat\'eriaux de
Strasbourg, CNRS (UMR 7504) and Universit\'e Louis Pasteur, 23 rue
du Loess, \New{BP 43, 67034 Strasbourg Cedex 2}, France.}

\author{F. Bardou}%
\affiliation{Institut de Physique et Chimie des Mat\'eriaux de
Strasbourg, CNRS (UMR 7504) and Universit\'e Louis Pasteur, 23 rue
du Loess, \New{BP 43, 67034 Strasbourg Cedex 2}, France.}

\begin{abstract}
The lognormal distribution describing, \eg exponentials of Gaussian
random
variables is one of the most common statistical distributions in physics.
It can exhibit features of broad distributions that imply qualitative
departure from the usual statistical scaling associated to narrow
distributions. Approximate formulae are derived for the
typical
sums of lognormal random variables. The validity of these formulae is
numerically checked and the physical consequences, \eg for
the current flowing through small tunnel junctions, are pointed out.
\end{abstract}


\maketitle

\section{Introduction: physics motivation}
\label{s1}
Most usual phenomena present a well defined average behaviour with
fluctuations around the average values. Such fluctuations
are described by narrow (or 'light-tailed') distributions like, \eg 
Gaussian or exponential distributions. Conversely, for other phenomena,
fluctuations themselves dictate the main features, while the average values
become either irrelevant or even non existent. 
Such fluctuations are described by broad (or 'heavy-tailed') distributions 
like, \eg distributions with power law tails generating 'L\'evy flights'.
After a long period in which the narrow distributions have had the 
quasi-monopoly of probability applications, it has been realized in 
the last fifteen years that broad distributions
arise in a number of physical systems \cite{BoG1990, SZF1995, PeS1999}.
 
Macroscopic physical quantities often appear as the sums $S_n$ 
of microscopic quantities $x_i$:
\begin{equation}
S_n = \sum_{i = 1}^n x_i,
\label{e1.1}
\end{equation}
where $x_1, x_2, \ldots, x_n$ are independent and identically distributed 
random variables.
The dependence of such sums $S_n$ with the number $n$ of terms epitomizes
the role of the broadness of probability distributions of $x_i$'s. One
intuitively expects the typical sum $\Vnt{S}$ to be given by:
\begin{equation}
\Vnt{S} \simeq n \average,
\label{e1.2}
\end{equation}
where $\average$ is the average value of $x$. The validity of \eq{e1.2}
is guaranteed at large $n$ by the law of large numbers. However, the law of 
large numbers is only valid for sufficiently narrow distributions. Indeed, for
broad distributions, the sums $S_n$ can strongly deviate from \eq{e1.2}. 
For instance, if the distribution of the $x_i$'s has a power law tail
(\cf L\'evy flights, \cite{BoG1990}), 
$\propto 1/x^{1+\alpha}$ with $0< \alpha <1$ ($\average = \infty$), then 
the typical sum of $n$ terms is not proportional to the 
number of terms but is given by:
\begin{equation}
\Vnt{S} \propto n^{1/\alpha}.
\label{e1.2bis}
\end{equation}
Physically, \eq{e1.2} (narrow distributions) and \eq{e1.2bis} (L\'evy flights)
correspond to different scaling behaviours.
For the L\'evy flight case, the violation of the law of large numbers occurs 
for any $n$. On the other hand, for other broad distributions like the 
lognormal treated hereafter, there is a violation of the law of large numbers 
only for finite, yet surprisingly large, $n$'s. 

These violations of the law 
of large numbers, whatever their extent, correspond physically to anomalous 
scaling behaviours as compared to those generated by narrow distributions.
This applies in particular to small tunnel junctions, such as the 
metal-insulator-metal junctions currently studied for
spin electronics \cite{MKW1995, MiT1995b}. It has indeed been shown, 
theoretically \cite{Bar1997} and experimentally \cite{DHB2000, Kel1999}, 
that these junctions tend to exhibit a broad distribution of tunnel 
currents that generates an anomalous scaling law: the typical
integrated current flowing through a junction is not proportional to the area of
the junction. This is more than just a theoretical issue since 
this deviation from the law of large numbers is most pronounced 
\cite{DHB2000, DRB2002}
for submicronic junction sizes relevant for spin electronics applications.

A similar issue is topical for the future development of metal oxide 
semiconductor field effect transistors (MOSFETs). Indeed, the downsizing of 
MOSFETs requires a reduction of the thickness of the gate oxide layer. 
This implies that tunnelling through the gate becomes non 
negligible\cite{CGD1999, MNO1998}, generating an unwanted current leakage. Moreover, 
as in metal-insulator-metal junctions, the large fluctuations of tunnel 
currents may give rise to serious irreproducibility issues. Our model 
permits a statistical description of tunnelling through non ideal barriers 
applying equally to metal-insulator-metal junctions and to MOSFET current 
leakages. Thus, anomalous scaling effects are expected to arise also in 
MOSFETs.

The current fluctuations in tunnel junctions are
well described by a lognormal probability density \cite{DBB1998, DHB2000}
\begin{equation}
f(x) = \LN (\mu, \sigma^2)(x) = 
\frac{1}{\sqrt{2 \pi \sigma^2} x}\exp\left[ - \frac{\left( \ln x
- \mu \right)^2}{2 \sigma^2} \right], x > 0
\label{e1.3}
\end{equation}
depending on two parameters, $\mu$ and $\sigma^2$. The
lognormal distribution presents at the same time features of a narrow
distribution, like the finiteness of all moments, and features of a broad
distribution, like a tail that can extend over several decades.  
It is actually one of the most common statistical distributions
and appears frequently, for instance, in biology \cite{LSA2001} and 
finance \cite{BoP2000} (for review see \cite{CrS1988, AiB1957}). In physics, it is 
often found in transport through
disordered systems such as wave propagation in random media (radar scattering, 
mobile phones,...)\cite{ScY1982, BAM1995}.
\New{A specially relevant example of the latter is transport through 1D 
disordered insulating wires for which the distribution of elementary resistances 
has been shown to be lognormal \cite{LaB1993}. This wire problem of random 
resistances in series is equivalent to the tunnel junction problem of 
random conductances in parallel \cite{RaR1989}. 
Thus, our results, 
initially motivated by sums of lognormal conductances in tunnel junctions, 
are also relevant for sums of lognormal resistances in wires.}

In this paper, our aim is to obtain analytical expressions 
for the dependence on the number $n$ of terms of the typical sums 
$\Vnt{S}$ of identically distributed lognormal random variables. The theory 
must treat the $n$ and $\sigma^2$ ranges relevant for applications. For
 tunnel junctions, both small $n \simeq 1$ corresponding to nanometric sized
junctions \cite{DBB1998} and large $n \simeq 10^{13}$ corresponding to 
millimetric sized junctions, and both small $\sigma^2 \simeq 0.1$ and large 
$\sigma^2 \simeq 10$ \cite{Dim2002, DTD2000, DDT2001, DHB2000} have been 
studied experimentally. For electromagnetic propagation in random media, 
$\sigma^2$ is typically in the range 2 to 10 \cite{BAM1995}.

There exist recent mathematical studies on sums of lognormal random variables 
\cite{BBG2002, BKL2002} that are motivated by glass physics (Random Energy 
Model). 
However, these studies apply to regimes of large $n$ and/or large $\sigma^2$ 
that do not correspond to those relevant for our problems. 
Our work concentrates on the deviation of the typical sum of a 
{\it moderate number} of lognormal terms with 
$\sigma^2 \lesssim 15$ from the asymptotic behaviour dictated by the law of 
large numbers. Thus, this paper and \cite{BBG2002, BKL2002} treat complementary 
$\left( n, \sigma^2 \right)$ ranges.

Section~\ref{s2} is a short review of the basic properties of lognormal
distributions, insisting on their broad character. 
Section~\ref{s3} presents qualitatively the sums of $n$
lognormal random variables. Section~\ref{s3.2} introduces the strategies 
used to estimate the typical sum $\Vnt{S}$. Section~\ref{Derivation}, the core 
of this work, derives approximate analytical expressions of $\Vnt{S}$ for 
different $\sigma^2$-ranges. 
Section~\ref{s4.1} discusses the range of validity of the obtained results. 
Section~\ref{s4.2} presents the striking scaling behaviour of the sample mean 
inverse.
Section~\ref{s5} contains a summarizing table and an overview of main results.

As the paper is written primarily for practitioners of quantum tunnelling, 
it reintroduces in simple terms the needed statistical notions about 
broad distributions. However, most of 
the paper is not specific to quantum tunnelling and its results may be
applied to any problem with sums of lognormal random variables.
The adequacy of the presented theory to describe experiments on
tunnel junctions is presented in \cite{DRB2002}.

\section{The lognormal distribution: simple properties and narrow {\it vs}
broad character}
\label{s2}
In this section, we present simple properties (genesis, characteristics, broad
character) of the lognormal distribution that will be used in the next sections.

Among many mechanisms that generate lognormal distributions \cite{CrS1988, AiB1957}, two 
of them are especially important in physics. 
In the first generation mechanism,
we consider $x$ as exponentially dependent on a Gaussian random variable $y$
with mean $\mu_y$ and variance $\sigma_y^2$:
\begin{equation}
x = x_0 e^{y/y_0}
\label{e2.1}
\end{equation}
where $x_0$ and $y_0$ are scale parameters for
$x$ and $y$, respectively.
The probability density of $y$ is:
\begin{equation}
\N (\mu_y, \sigma_y^2) (y)= 
\frac{1}{\sqrt{2 \pi \sigma_y^2} } \exp \left[ - \frac{\left( y - \mu_y 
\right)^2}{2 \sigma_y^2}\right].
\label{e2.2}
\end{equation}
The probability density of $x$, $f(x) = \N (\mu_y, \sigma_y^2) (y) \diff{y} / \diff{x}$ is a 
lognormal density $\LN \left(\mu, \sigma^2\right)\left(x\right)$, as in 
\eq{e1.3}, with parameters:
\begin{subequations}
\label{e2.3}
\begin{equation}
\mu = \frac{\mu_y}{y_0} + \ln{x_0},
\label{e2.3a}
\end{equation}
\begin{equation}
\sigma^2 = \left( \sigma_y / y_0 \right)^2.
\label{e2.3b}
\end{equation}
\end{subequations}
A typical example of such a generation mechanism is provided by tunnel 
junctions. Indeed, the exponential current dependence on the potential 
barrier parameters operates as a kind of 'fluctuation amplifier' by 
non-linearly transforming small Gaussian fluctuations of the parameters into 
qualitatively large current fluctuations. This implies, as seen above, 
lognormal distribution of tunnel currents \cite{DRB2002}.

In the second generation mechanism, we consider the product 
$x_n = \prod_{i = 1}^n{y_i}$ of $n$ identically distributed random variables 
$y_1, \cdots, y_n$. If $\mu'$ and $\sigma'$ are the mean and the standard 
deviation of $\ln{y_i}$, not necessarily Gaussians, then
\begin{equation}
\ln{x_n} = \sum_{i = 1}^n{\ln{y_i}}
\label{e2.4}
\end{equation}
tends, at large $n$, to a Gaussian random variable of mean $n \mu'$ and variance
$n \sigma{'}^2$, according to the central limit theorem. 
Hence, using \eqs{e2.3a}{e2.3b} with $x_0 = y_0 = 1$,
 $x_n$ is lognormally distributed with parameters 
$\mu = n \mu'$ and $\sigma^2 = n \sigma{'}^2$. \New{For a better approximation 
at finite $n$, see \cite{Red1990}.}

The lognormal distribution given by \eq{e1.3} has the following characteristics.

The two parameters $\mu$ and $\sigma^2$ are, according to \eq{e2.3a} 
and \eq{e2.3b} with $x_0=y_0=1$, 
the mean and the variance of the Gaussian random variable 
$\ln x$. The parameter $\mu$ is a {\it scale parameter}. Indeed, if $x$ 
is distributed according to $\LN(\mu,\sigma^2)(x)$, then $x'=\alpha x$ is 
distributed according to $\LN(\mu'= \mu+\ln \alpha,{\sigma '}^2 
= \sigma^2)(x')$,
as can be seen from \eq{e2.1}, \eq{e2.3a} and \eq{e2.3b}. Thus, one can always
take $\mu =0$ using a suitable choice of units. On the other hand, $\sigma^2$ 
is the {\it shape parameter} of the lognormal distribution.

The {\it typical value} $\xt$, corresponding to the maximum of the 
distribution, is
\begin{equation}
x^\mathrm{t} = e^{\mu-\sigma^2}.
\label{e2.5}
\end{equation}
The {\it median}, $\xm$, such that 
$\int_0^\xm f(x)\diff x = \int_\xm^\infty f(x)\diff x = 1/2$, is
\begin{equation}
\xm = e^\mu.
\label{e2.6}
\end{equation}
The {\it average}, $\average$, and the {\it variance}, 
$\var{x}  \defin \langle x^2 \rangle -\average^2$, are
\begin{eqnarray}
\label{e2.7m}
\average & = & e^{\mu+\sigma^2/2},  \\
\var{x} & = & e^{2\mu+ \sigma^2} \left( e^{\sigma^2} -1 \right).
\label{e2.7s}
\end{eqnarray}
The {\it coefficient of variation},  
$C \defin \sqrt{\var{x}}/\average$, which characterizes the relative dispersion of
the distribution, is thus
\begin{equation}
C = \sqrt{e^{\sigma^2} -1}.
\label{e2.8}
\end{equation}
Note that $\mu$ does not appear in $C$, as expected for a scale parameter.

\begin{figure}
\includegraphics[scale=0.33,angle=-90]{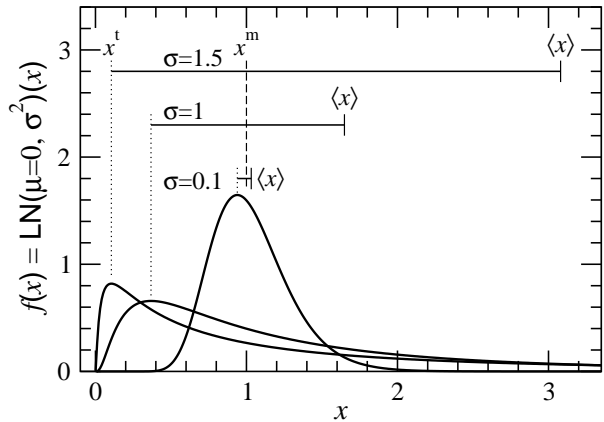}
\caption{Examples of lognormal distributions $\LN\left( \mu, \sigma^2 \right)
\left( x\right)$ with $\mu = 0$ and $\sigma = $ 0.1, 1 and 1.5. When $\sigma$
increases, the typical values $\xt$, indicated by the dotted lines, and the 
means $\average$ move rapidly away from the constant median $\xm$, indicated by 
the broken line, in opposite directions.}
\label{fig2.1}
\end{figure}
\Fig{fig2.1} shows examples of lognormal distributions with scale parameter
$\mu = 0$ and different shape parameters. For small
$\sigma^2$, the lognormal distribution is
narrow (rapidly decaying tail) and can be approximated by a Gaussian 
distribution (see \appen{a1}). When $\sigma^2$ increases, the lognormal 
distribution rapidly becomes broad (tail extending to values much larger than 
the typical value). In particular, the typical value $\xt$ and the mean 
$\average$ move in opposite directions away from the median 
$\xm$ which is 1 for all $\sigma^2$. The strong
$\sigma^2$-dependence of the broadness is quantitatively given by
the coefficient of variation, \eq{e2.8}. 

Another way of characterizing 
the broadness of a distribution, is to define an interval containing a 
certain percentage of the probability. For the Gaussian distribution 
$\N\left( \mu, \sigma^2 \right)$, 68\% of the probability is contained
in the interval $\left[ \mu - \sigma, \mu + \sigma\right]$ whereas for 
the lognormal distribution $\LN\left( \mu, \sigma^2 \right)$, the same 
probability is contained within $\left[ \xm / e^\sigma, \xm \times 
e^\sigma\right]$.
The extension of this interval depends linearly on $\sigma$ for the
Gaussian and exponentially for the lognormal.

Moreover, the weighted distribution $x f\left(x\right)$, giving the 
distribution of the contribution to the mean, is peaked on the median $\xm$. 
In the vicinity
of $\xm$ one has\cite{MoS1983}:
\begin{equation}
f(x) = \frac{1}{\sqrt{2\pi \sigma^2} x} ~\mathrm{for}~ e^{\mu-\sqrt{2} \sigma} \ll x 
\ll e^{\mu+\sqrt{2} \sigma}.
\label{e2.9bis}
\end{equation}
Thus, $f\left(x\right)$ behaves as a distribution that is extremely broad 
($1 / x$ is not even normalizable) in an $x$-interval whose size increases 
exponentially fast with $\sigma$ and that is smoothly truncated outside this 
interval.

Three different regimes of broadness can be defined using the peculiar 
dependence of the probability peak height $f\left( \xt \right)$
on $\sigma^2$.
Indeed, the use of \eq{e1.3} and \eq{e2.5} yields:
\begin{equation}
f(\xt) = \frac{e^{\sigma^2/2}}{\sqrt{2\pi} e^\mu \sigma}.
\label{e2.9}
\end{equation}

For $\sigma^2 \ll 1$, one has $f(\xt) \propto 1 / \sigma$ and thus 
$f(\xt) \propto 1 / \sqrt{\var{x}}$ as $\sqrt{\var{x}} \propto \sigma$ (see \eq{e2.7s}). This inverse
proportionality between peak height $f(\xt)$ and peak width $\sqrt{\var{x}}$ is the
usual behaviour for a narrow distribution that concentrates most of the
probability into the peak.

When the shape parameter $\sigma^2$ increases, still keeping $\sigma^2 \leq 1$, 
$f(\xt)$ is no longer inversely proportional to $\sqrt{\var{x}}$, however it still
decreases, as expected for a distribution that becomes broader and thus less
peaked (see, in \fig{fig2.1}, the difference between $\sigma = 0.1$ and 
$\sigma = 1$). 

On the contrary, when $\sigma^2 > 1$, {\it the peak height increases
with $\sigma^2$} even though the distribution becomes broader (see, 
in \fig{fig2.1}, the difference between $\sigma = 1$ and $\sigma = 1.5$).
This is more unusual. The behaviour of the peak can be understood from the 
genesis of the lognormal variable $x = e^y$ with $y$ distributed as 
$\N \left( \mu_y = \mu, \sigma_{y}^2 = \sigma^2 \right)\left( y \right)$. 
When $\sigma^2$ 
becomes larger, the probability to draw $y$ values much smaller than $\mu$
increases, yielding many $x$ values much smaller than $e^{\mu}$, all packed
close to $0$. This creates a narrow and high peak for $f(x)$.

This non monotonous variation of the probability peak $f(\xt)$ with the 
shape parameter $\sigma^2$ with a minimum in $\sigma^2 = 1$, incites to 
consider three qualitative classes of lognormal distributions, that will
be used in the next sections. The class $\sigma^2 \ll 1$ corresponds to the narrow
lognormal distributions that are approximately Gaussian. The class 
$\sigma^2 \lesssim 1$ contains the moderately broad lognormal 
distributions that may 
deviate significantly from Gaussians, yet retaining some features of narrow 
distributions. The class $\sigma^2 \gg 1$ contains the very broad lognormal
distributions.

\section{Qualitative behaviour of the typical sum of lognormal random variables}
\label{s3}
In this section we explain the qualitative behaviour of the 
typical sum of lognormal random variables by relating it to the behaviours of
narrow and broad distributions. 

\begin{figure}
\includegraphics[scale=1.00,angle=0]{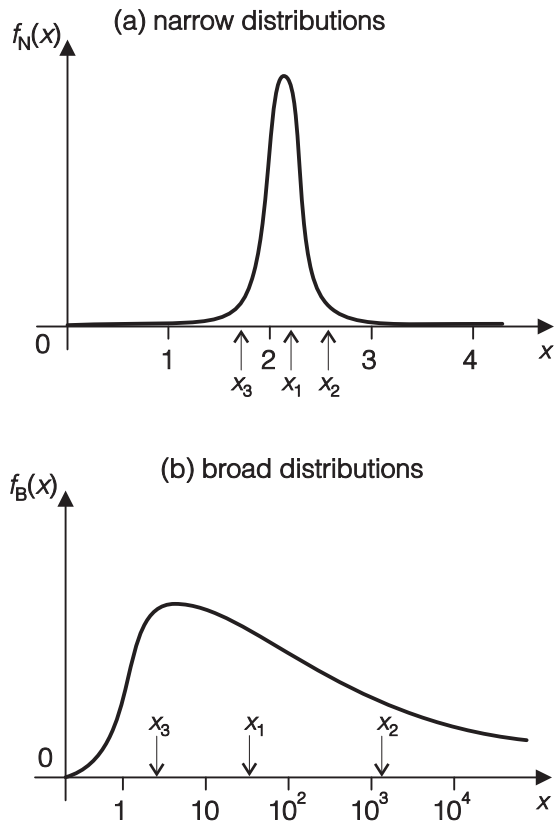}
\caption{Narrow \vs broad distributions. (a) A narrow distribution $\fN(x)$
presents a well defined peak and light tails. In a set $\{x_1, \ldots, x_n\}$
of $n$ random numbers drawn from $\fN(x)$, no number is dominant.
(b) A broad distribution $\fB(x)$ presents a long tail extending over 
several decades (note the logarithmic $x$-scale). In a set 
$\{x_1, \ldots, x_n\}$
of $n$ random numbers drawn from $\fB(x)$, one number is clearly dominant.}
\label{fig3.1}
\end{figure}

Consider first a narrow distribution $\fN(x)$ presenting a well defined 
narrow peak concentrating most of the probability in the vicinity of the 
mean $\average$ and with light tails decaying sufficiently 
rapidly away from the peak (\fig{fig3.1}a).
Draw, for example, three random numbers $x_1$, $x_2$ and $x_3$ according 
to the distribution $\fN(x)$. If $\fN(x)$ is sufficiently narrow, then 
$x_1$, $x_2$ and $x_3$ will all be approximately equal to each other and to the 
mean $\average$ and thus,
\begin{equation}
S_3 = x_1 + x_2 + x_3 \simeq 3 x_{1, \, 2 \, \mathrm{or} \, 3} 
\simeq 3 \average .
\label{e3.1.1}
\end{equation}
Note that no single term $x_i$ dominates the sum $S_3$. More generally, 
the sum of $n$ terms will be close, even for small $n$'s, to the large 
$n$ expression given by the law of large numbers:
\begin{equation}
S_n \simeq n \average.
\label{e3.1.2}
\end{equation}

Consider now a broad distribution $\fB(x)$ whose probability spreads throughout
a long tail extending over several decades (\fig{fig3.1}b; note the 
logarithmic $x$-scale) instead of being concentrated into a peak. Drawing three
random numbers according to $\fB(x)$, it is very likely that one of these 
numbers, for example $x_2$, will be large enough, compared to the other ones,
to dominate the sum $S_3$:
\begin{equation}
S_3 = x_1 + x_2 + x_3 \simeq \max(x_1,x_2,x_3) = x_2 .
\label{e3.1.3}
\end{equation}
More generally, the largest term $M_n$,
\begin{equation}
M_n \defin \max(x_1,\ldots,x_n),
\label{e3.1.4bis}
\end{equation}
will dominate the sum of $n$ terms:
\begin{equation}
S_n \simeq M_n.
\label{e3.1.4}
\end{equation}

Under these premises, what is the order of magnitude of $S_n$? To approximately estimate it, 
one can divide the interval $[0;\infty)$ of possible values of $x$
\footnote {We assume for simplicity that $x$ is positive.} into $n$ intervals 
$[a_1=0;a_2)$, $[a_2;a_3)$, ..., $[a_n;a_{n+1}=\infty)$ corresponding
to a probability of $1/n$:
\begin{equation}
\frac{1}{n} = \int_{a_j}^{a_{j+1}} \fB(x) \;\diff x \, .
\label{e3.1.6}
\end{equation}
Intuitively, there is typically one random number $x_i$ in each interval 
$[a_j;a_{j+1})$. The largest number $M_n$ is thus very likely to lie in 
the rightmost interval $[a_n;\infty)$. The most probable number in this 
interval is $a_n$ (we assume that $\fB(x)$ is decreasing at large $x$).
Thus, applying \eq{e3.1.4} the sum $S_n$ is approximately given by:
\begin{equation}
S_n \simeq a_n \ \mathrm{with} \ \frac{1}{n} = \int_{a_n}^{\infty} \fB(x) \; \diff x.
\label{e3.1.7}
\end{equation}

As a specific application, consider for example
a Pareto distribution $\fP(x)$ with infinite mean,
\begin{equation}
\fP(x) \defin \frac{\alpha x_0^\alpha}{x^{1+\alpha}}, \quad \mathrm{for} \quad
x \geq x_0 \quad \mathrm{and} \quad \mathrm{with} \quad 0<\alpha<1 .
\label{e3.1.8}
\end{equation}
In this case, the sum $S_n$ is called a 'L\'evy flight'. The relation
~(\ref{e3.1.7}) yields\footnote{A rigorous derivation of $\Vnt{M}$ based on 
order statistics gives 
$\Vnt{M} = x_0 \left( \frac{1+\alpha n}{1+\alpha} \right)^{1/\alpha}$
(see, \eg eq.~(4.32) in \cite{BBA2002}). This expression is close to \eq{e3.1.9},
which consolidates the intuitive reasoning based on \eq{e3.1.7} to derive
$\Vnt{M}$.
} 
$\Mnt \simeq x_0 n^{1/\alpha}$ and thus, using \eq{e3.1.4},
\begin{equation}
\Snt \simeq x_0 n^{1/\alpha} .
\label{e3.1.9}
\end{equation}
Note that, as $\alpha < 1$, the average value is infinite and thus the law 
of large numbers does not apply here.

\begin{figure}
\includegraphics[scale=0.33,angle=-90]{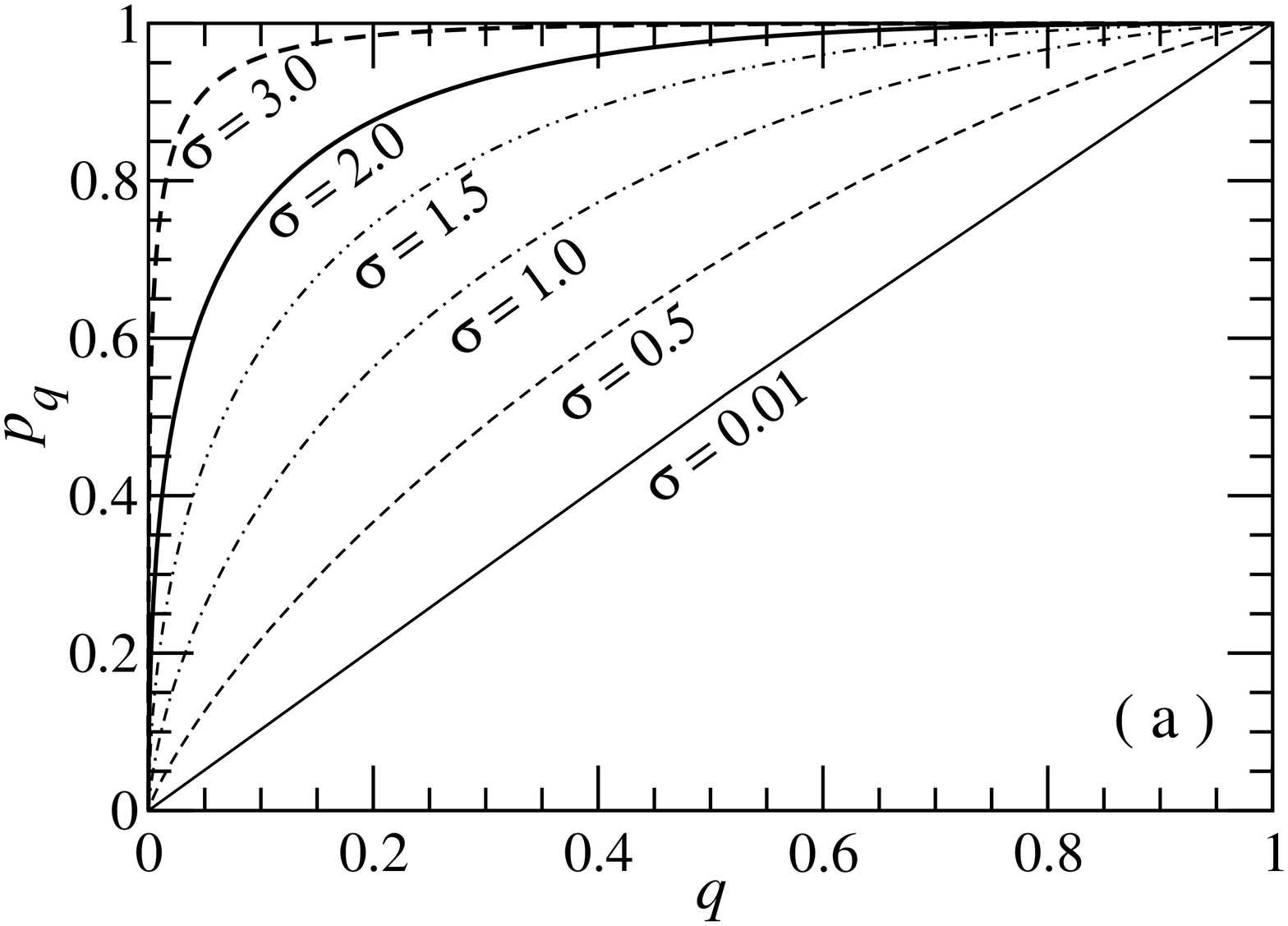}
\includegraphics[scale=0.33,angle=-90]{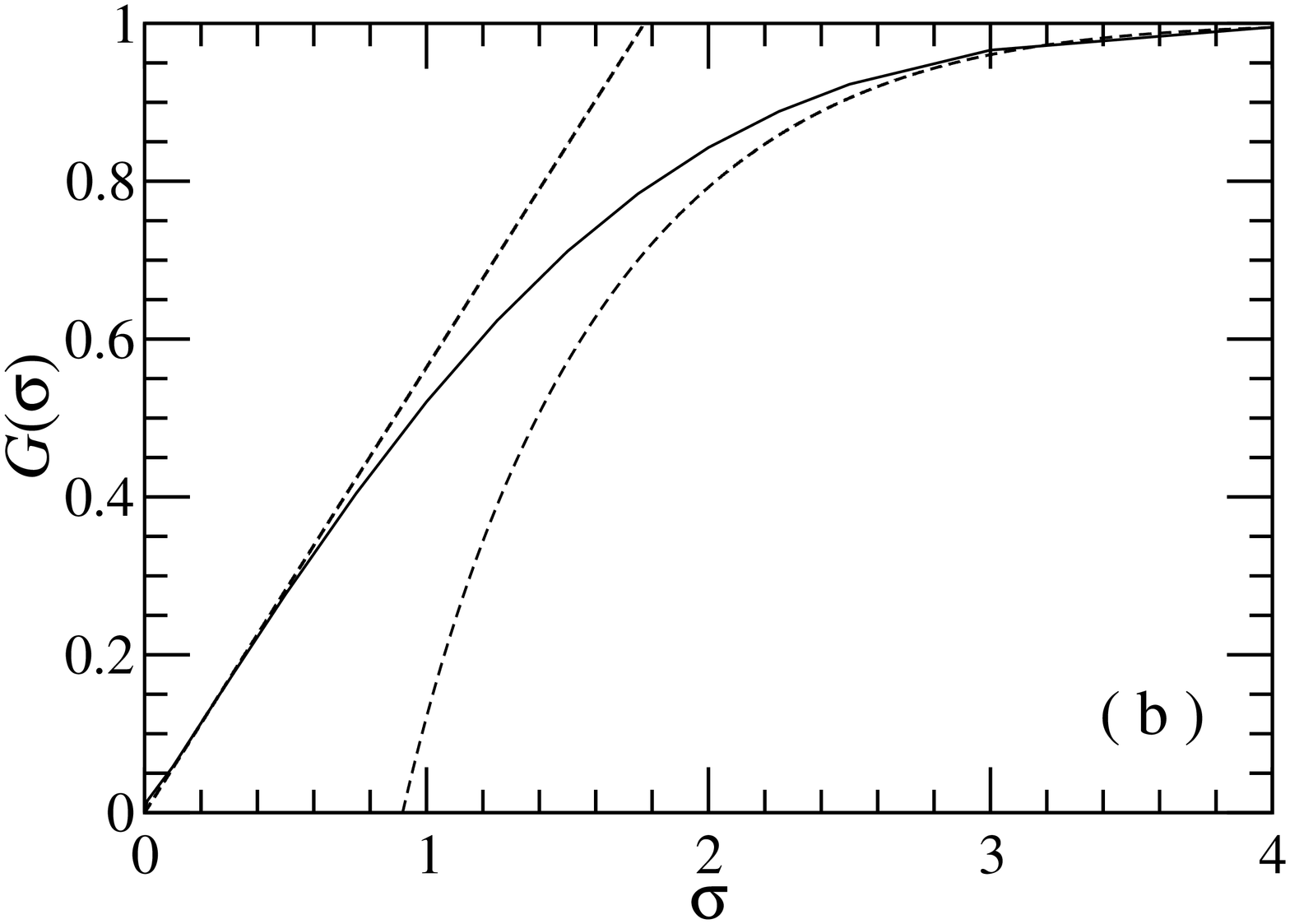}
\caption{Heterogeneity of the terms of lognormal sums. a) Average proportion 
$p_q$ of the average physical quantity $\average$ carried by 
the average proportion $q$ of the statistical sample. For narrow lognormal 
distributions $\left( \sigma^2 \ll 1 \right)$, all terms equally contribute to 
the sums $\left( p_q \simeq q \right)$. For broad lognormal distributions, 
a small proportion of the terms provide the major contribution to the sums 
$\left( p_q \gg q \mathrm{\ for \ } q \ll 1\right)$. b) Gini coefficient 
giving a quantitative measure of the heterogeneity.}
\label{fig3.2}
\end{figure}

The fact that the sum $S_n$ of $n$ terms increases typically faster in 
\eq{e3.1.9} than the number $n$ of terms is in contrast with the law 
of large numbers. This 'anomalous' behaviour can be intuitively explained
(see also \fig{fig3.2} in \cite{DRB2002} for a complementary approach).
Each draw of a new random number from a broad distribution $\fB(x)$ gives the
opportunity to obtain a large number, very far in the tail, that will 
dominate the sum $S_n$ and will push it towards significantly larger values. 
Conversely, for narrow distributions $\fN(x)$, the typical largest term 
$\Vnt{M}$ increases very slowly with the number of terms
({\it e.g.}, as $\sqrt{\ln n}$ for a Gaussian distribution and as $\ln n$ for an
exponential distribution; see, \eg \cite{EKM1997}), whilst the typical sum 
$\Vnt{S}$ increases linearly with $n$ and thus $\Vnt{S} \gg \Vnt{M}$. 

The question that arises now is whether the sum of lognormal random variables
behaves like a narrow or like a broad distribution.
On one hand, the lognormal distribution has finite moments, like a narrow 
distribution. Therefore, the law of large numbers must apply at least 
for an asymptotically large number of terms: 
$S_n \oper{\to}{n\to \infty} n \average$. 
On the other hand, if $\sigma^2$ is sufficiently large, the lognormal tail
extends over several decades, as for a broad distribution (see \Section{s2}).
Therefore, the sum of $n$ terms is expected to be dominated by a small 
number of terms, if $n$ is not too large\footnote{
This is distinct from the subexponential property.
The subexponentiality of the lognormal distribution \cite{Asm2000} ensures that 
{\it asymptotically large} sums $S_n$ are dominated by the largest term, 
for any $n$. Here on the contrary, we are interested 
in the domination of the {\it typical} sum, which is by definition {\it not} asymptotically 
large, by the largest term, a property that is only valid for a limited 
$n$ range.
}. 

The domination of the sum by the largest terms can be quantitatively estimated
by computing the relative contribution $p_q$ to the mean by the proportion
$q$ of statistical samples with values larger than some $x_q$\footnote{
	The expressions of $p_q$ and $q$ given by \eq{e.quantile.a}
	and \eq{e.quantile.b} are meaningful for very large statistical 
	samples, as they correspond to average quantities. For small samples, 
	statistically, $p_q$ and $q$ might deviate significantly from these 
	expressions.
	}\footnote{
	For tunnel junctions, the plot $p_q$ \vs $q$ gives a measure of 
	the inhomogeneity 
	corresponding to the so-called 'hot spots':
	$p_q$ is the proportion of the average current carried by the 
	proportion $q$ of the junction area with currents larger than $x_q$.
	}
\begin{equation}
p_q \defin \int_{x_q}^\infty x' f(x') \diff x' /\average,
\label{e.quantile.a}
\end{equation}
\begin{equation}
q \defin \int_{x_q}^\infty f(x') \diff x'.
\label{e.quantile.b}
\end{equation}
\Fig{fig3.2}\New{a} shows a plot
of $p_q$ \vs $q$ for various $\sigma$'s. \New{Note that the curve 
$\left( 1-q, 1-p_q \right)$ is called a Lorenz plot in the economics 
community when studying the distribution of incomes (see, \eg \cite{DrY2001}).}
For small $\sigma$'s 
($\sigma \lesssim 0.25$), one has $p_q \simeq q$ for all $q$: all terms 
$x_i$ equally contribute to the sum $S_n$. This is the usual behaviour 
of a narrow distribution. For larger $\sigma$'s, one has $p_q \gg q$ for 
$q \ll 1$: only a small number of terms contribute significantly to the
sum $S_n$. This is the usual behaviour of a broad distribution. Monte Carlo 
simulations of tunnelling through MOSFET gates yield $p_q$ \vs $q$ curves 
that are strikingly similar to \fig{fig3.2}\New{a} (see figure 11 of \cite{CGD1999}). 
Indeed, the parameters used in \cite{CGD1999} correspond to a barrier 
thickness standard deviation of $\sigma_d = 0.18\ \mathrm{nm}$, a barrier 
penetration length $\lambda \simeq 7.8 \times 10^{-2}\ \mathrm{nm}$ which 
gives $\sigma = \sigma_d / \lambda \simeq 2.3$ (see \cite{DHB2000} or 
\cite{DRB2002} for the derivation of $\sigma = \sigma_d / \lambda$). For 
this $\sigma$, figure 11 of \cite{CGD1999} fits our $p_q$ \vs $q$ without 
any adjustable parameter. 

\New{As in economics, the information contained in \fig{fig3.2}a can be 
summarized by the Gini coefficient $G$ represented in \fig{fig3.2}b:
\begin{equation}
G \defin 2 \int_{0}^1 \left( p_q - q \right) \diff q,
\label{e.Gini.1}
\end{equation}
giving a quantitative measure of the heterogeneity of the contribution of 
the terms to the sum. In the lognormal case this expression becomes: 
$G\left( \sigma \right) = 
1 - 2 \int_{-\infty}^{\infty} \N\left( 0, 1 \right)\left( u \right) 
\Phi\left( u - \sigma \right) \diff u$, where $\N\left( 0, 1 \right)\left( u 
\right)$ is the normal distribution (\eq{e2.2}) and 
$\Phi\left( u \right) \defin \int_{-\infty}^{u} \N\left( 0, 1 \right)\left( u' \right) 
\diff u'$ the corresponding distribution function. The solid line 
in \fig{fig3.2}b represents $G\left( \sigma \right)$ for 
various $\sigma$'s. As expected, $G\left( \sigma \right)$ varies from 0 
when $\sigma = 0$, which means that all terms of a narrow lognormal 
distribution equally contribute to the sums, to 1 when $\sigma \to \infty$, 
which means that only a small proportion of the terms of a broad lognormal 
distribution contributes significantly to the sums. The broken lines in 
\fig{fig3.2}b represent analytically derived asymptotic approximations of 
$G\left( \sigma \right)$:
\begin{subequations}
\begin{equation}
\sigma \ll 1 : G\left(\sigma\right) \simeq \frac{\sigma}{\sqrt{\pi}}
\label{e.Gini.2.1}
\end{equation}
\begin{equation}
\sigma \gg 1 : G\left(\sigma\right) \simeq 1 - \frac{2 e^{-\sigma^2 / 4}}
{\sqrt{\pi}\sigma}
\label{e.Gini.2.2}
\end{equation}
\end{subequations}
(Our derivations of these formulae, which are not explicitely shown 
here, are based on usual expansion 
techniques).}

In summary, if $\sigma^2$ is small, the sum of $n$ lognormal terms is expected to 
behave like sums of narrowly distributed random variables, for any $n$.
Conversely, if $\sigma^2$ is sufficiently large, the sum of $n$ lognormal 
terms is expected to behave, at small $n$, like sums of broadly 
distributed random variables and, at large $n$, like sums of narrowly 
distributed random variables (law of large numbers). Before converging to the 
law of large numbers asymptotics, the typical sum may deviate strongly 
from this law. Moreover, if this convergence is 
slow enough, physically relevant problems may lie in the non converged 
regime. This is indeed the case of submicronic tunnel junctions
\cite{DHB2000}.

\section{Strategies for estimating the typical sum}
\label{s3.2}
In this section, we discuss strategies 
for obtaining the typical sum $\Vnt{S}$ of lognormal random variables
 depending on 
the value of the shape parameter $\sigma^2$. 

\begin{figure}
\includegraphics[scale=0.85,angle=0]{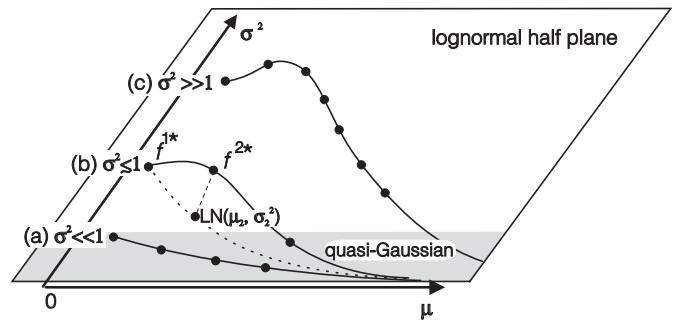}
\caption{{\it Schematic representation of the trajectory of $f^{n*}$ in the
space of distributions.} The set of lognormal distributions
corresponds to the open half-plane $\left( \mu \in \mathbb{R}, \sigma^2 > 0 \right)$.
The infinite dimension space of probability distributions is schematically
represented here in three dimensions. In the region $\left( \mu \in \mathbb{R},
\sigma^2 \ll 1 \right)$ (shaded area), lognormal distributions are
quasi-Gaussian. (a) Narrow lognormal distributions $f^{1*} (\sigma^2 \ll 1)$ are
quasi-Gaussian and, thus, the trajectory of $f^{n*}$ starts and ends up in the
close vicinity of the line $\left( \mu \in \mathbb{R}, \sigma^2 = 0 \right)$. (b) For
moderately broad lognormal distributions ($\sigma^2 \lesssim 1$), the trajectory
of $f^{n*}$ starts in the lognormal half-plane, not too far away from the
quasi-Gaussian region, that is reached for asymptotically large $n$. Thus
$f^{n*}$ is conjectured to lie, for any $n$, close to the lognormal half-plane:
$f^{n*} \simeq \LN(\mu_n, \sigma_n^2)$. (c) Very broad lognormal distributions
$f^{1*} (\sigma^2 \gg 1)$ lie far away from the quasi-Gaussian region. Thus, there
is a long way before  $f^{n*}$ enters the quasi-Gaussian region and $f^{n*}$
has the possibility to come significantly out of the lognormal half-plane for
intermediate values of $n$.}
\label{fig3.3}
\end{figure}

By definition $\Vnt{S}$ is the
peak position of the distribution of $S_n$. Moreover, the latter is
 the $n$-fold
convolution of $f(x)$ and is denoted as $f^{n*}(S_n)$.
As no exact analytical expression is known for $f^{n*}$ when $f$ is lognormal,
one will turn to approximation strategies. These strategies can be derived
from the schematic representation, in the space of distributions, of
the trajectory followed by $f^{n*}$ with increasing $n$ (\fig{fig3.3}).
The set of lognormal distributions can be represented by
an open half-plane ($\mu$, $\sigma^2$) with $\mu \in (-\infty; \infty)$ and
$\sigma^2 \in (0; \infty)$. In this half-plane, the shaded region with 
$\sigma^2 \ll 1$ corresponds to quasi-Gaussian lognormal distributions (see 
\appen{a1}). The whole lognormal half-plane is embedded in the infinite
dimension space of probability distributions, which is schematically represented
in \fig{fig3.3} as a three dimension space. 

\bigskip
The starting point $f^{1*}$ and the asymptotic behaviour $f^{n*}$ with 
$n \to \infty$ of the $f^{n*}$ trajectory are trivially known for any $\sigma^2$. 
 Indeed, $f^{1*} = f = \LN(\mu, \sigma^2)$ lies exactly in the lognormal 
 half-plane. Moreover, the finiteness 
of the moments of the lognormal distribution
$f^{1*} = f$ implies the applicability of the central limit theorem:
\begin{equation}
f^{n*}(S_n) \oper{\to}{n\to \infty} N\left(n \average, n \sigma^2\right)(S_n),
\label{e3.2.2}
\end{equation}
where $N\left(n \average, n \sigma^2\right)(S_n)$ is narrow since its 
coefficient of 
variation $\sqrt{n\sigma^2}/n\average \propto 1/\sqrt{n}$ 
tends to zero. As narrow Gaussian distributions are quasi-lognormal, as shown 
in \appen{a1},  $f^{n*}$ lies close to the 
quasi-Gaussian region of the lognormal half-plane. 

\bigskip
For intermediate $n$, on the contrary, the trajectory of $f^{n*}$ strongly 
depends on the broadness of the 
initial lognormal distribution $f^{1*}$ and 
three different cases can be distinguished.

For narrow lognormal distributions ($\sigma^2 \ll 1$), both the starting point
$f^{1*}$ and the end point $f^{n*}$ for $n = \infty$ belong to the 
quasi-Gaussian region. 
Therefore one can assume that $f^{n*}$ is quasi-Gaussian for any $n$, which 
gives immediately the typical sums $\Vnt{S}$ (see \Section{s3.3}).

For moderately broad lognormal distributions ($\sigma^2 \lesssim 1$), 
$f^{n*}$ does not start too far away from the quasi-Gaussian region 
that is reached at large $n$. Hence, one can assume that $f^{n*}$ 
remains close to the lognormal half-plane\footnote{The family of lognormal 
distributions is
not closed under convolution. Thus, it is clear that $f^{n*}$ {\it is not
exactly lognormal}.} in between $n = 1$ and $n = \infty$. 
Thus, the approximation strategy 
will consist in finding a lognormal distribution $\LN(\mu_n, \sigma_n^2)$ 
(see broken line in \fig{fig3.3}) that closely
approximates $f^{n*}$ (see Section~\ref{s3.4}).

For very broad lognormal distributions $(\sigma^2 \gg 1)$, $f^{n*}$ starts far 
away from the quasi-Gaussian region 
that is reached at large $n$. Hence, $f^{n*}$ may significantly come out of
the lognormal half-plane. In this case, the approximation strategy is
dictated by the fact that sums $S_n$ are dominated by the largest terms 
(see Section~\ref{s3.5}).

\section{Derivation of the typical sums of lognormal random variables}
\label{Derivation}
In this section we apply the strategies discussed above in order to derive 
approximate analytical 
expressions of $\Vnt{S}$ for different ranges of $\sigma^2$.

\subsection{Case of narrow lognormal distributions}
\label{s3.3}
We consider here the case $\sigma^2 \ll 1$ of narrow lognormal distributions. 
As seen in \appen{a1}, a narrow lognormal distribution is well approximated by a
normal distribution:
\begin{equation}
\sigma^2 \ll 1 : \LN\left( \mu, \sigma^2 \right) \simeq N\left( e^{\mu}, \left(
\sigma e^{\mu} \right)^2 \right) \nonumber
\end{equation}
Consequently, the typical sum $\Vnt{S}$ is simply given by:
\begin{equation}
\sigma^2 \ll 1 : \Vnt{S} \simeq n e^{\mu}
\label{e3.3.7}
\end{equation}
as in the Gaussian case, for any number of terms. Note that the law of large
numbers asymptotics $\Vnt{S} \oper{\to}{} n \average = n e^{\mu +
\sigma^2/2}$, close to \eq{e3.3.7} for $\sigma^2 \ll 1$, is applicable
here even for a small number of terms.

\subsection{Case of moderately broad lognormal distributions}
\label{s3.4}
We consider here the case $\sigma^2 \lesssim 1$ of moderately broad 
lognormal distributions  that already allows considerable deviation from the
Gaussian behaviour obtained for $\sigma^2 \ll 1$ (see Section~\ref{s3.3}). The
distribution $f^{n*}$ of $S_n$ is now conjectured to be close to a lognormal
distribution:
\begin{equation}
\sigma^2 \lesssim 1 : f^{n*}\left( S_n \right) \simeq \LN\left( \mu_n, \sigma_n^2
\right)\left( S_n \right)
\label{e3.4.1}
\end{equation}
Two equations characterizing $f^{n*}$ are needed to determine the two unknown
parameters $\mu_n$ and $\sigma_n^2$.

The cumulants provide such exact relationships on $f^{n*}$. In particular, the
first two cumulants\footnote{The choice of the first two cumulants results from
a compromise. We are looking for the typical value $\Vnt{S} = e^{\mu_n -
\sigma_n^2}$ of $f^{n*} \simeq \LN\left( \mu_n, \sigma_n^2 \right)$ which is
smaller than both $\langle S_n \rangle = e^{\mu_n + \sigma_n^2 / 2}$ and
$\langle S_n^2 \rangle^{1/2} = e^{\mu_n + \sigma_n^2}$. Therefore, the
first two cumulants, $\langle S_n \rangle$ and $\var{S_n}$
(involving $\langle S_n^2 \rangle$), give informations on two quantities larger
than $\Vnt{S}$. It would have been preferable to use one quantity larger and another
one smaller than $\Vnt{S}$, but this is not possible with cumulants. Hence, the
least bad choice is to take the cumulants that involve the quantities $\langle
S_n^k \rangle^{1/k}$ that are the least distant from $\Vnt{S}$, \ie the
cumulants of lowest order: $\langle S_n \rangle$ and $\var{S_n}$.
Similar uses of cumulants to find approximations of the $n$-fold convolution of
lognormal distributions have also been proposed in the context of radar scattering
\cite{Fen1960} and mobile phone electromagnetic propagation \cite{ScY1982}.}, 
$\langle S_n \rangle$ and $\var{S_n}$ obey:
\begin{subequations}
\label{e3.4.2}
\begin{equation}
\langle S_n \rangle = n \average
\label{e3.4.2a}
\end{equation}
\begin{equation}
\var{ S_n } = n \var{x}.
\label{e3.4.2b}
\end{equation}
\end{subequations}
These equations imply
\begin{equation}
C_n^2 = \frac{C^2}{n},
\label{e3.4.4}
\end{equation}
where $C_n \defin \left[ \var{S_n}\right]^{1/2}/\langle S_n \rangle$ is 
the coefficient of variation of $S_n$%
\footnote{Physically, \eq{e3.4.4} corresponds to the
usual decrease as $1 / \sqrt{n}$ of the relative fluctuations $C_n$ with the
size $n$ of the statistical sample.}%
. As $f$ is lognormal and $\fnstar$
is approximately lognormal, one has $C^2=e^{\sigma^2}-1$ and
$C_n^2=e^{\sigma_n^2}-1$ (see \eq{e2.8}). 
Then, using \eq{e3.4.4}, we obtain 
\begin{equation}
\sigma_n^2 = \ln \left(1+ \frac{e^{\sigma^2}-1}{n} \right)
	= \ln \left(1+ \frac{C^2}{n} \right).
\label{e3.4.5}
\end{equation}
At last, we derive $\mu_n$ by developing \eq{e3.4.2a} using \eq{e2.7m}:
\begin{equation}
e^{\mu_n + \sigma_n^2/2} = n e^{\mu+\sigma^2/2}.
\end{equation}
Thus, thanks to \eq{e3.4.5}, one has:
\begin{eqnarray}
\mu_n &=& \mu + \frac{\sigma^2}{2} + 
		\ln\left( \frac{n}{\sqrt{1+\frac{e^{\sigma^2}-1}{n}}}\right) 
		\nonumber \\
	&=& \ln\left( n \average \right) - \frac{1}{2} \ln \left( 1+ 
						\frac{C^2}{n} \right) 
\label{e3.4.7}
\end{eqnarray}

In the remainder of this section, we will examine the consequences of
\eq{e3.4.5} and \eq{e3.4.7} on the typical sum 
$\Vnt{S}$, on the height of the peak of $f^{n*}$ and on the convergence 
of $f^{n*}$ to a Gaussian.

The typical sum $\Vnt{S}$ derives from \eq{e3.4.5}, \eq{e3.4.7} 
and \eq{e2.5}:
\begin{equation}
\sigma^2 \lesssim 1: \quad \Vnt{S} \simeq n \average \frac{1}{\left( 1 + \frac{C^2}{n} \right)^{3/2}} .
\label{e3.4.9}
\end{equation}
The typical sum $\Snt$ appears as the product of the usual law of large numbers,
$n\average$, and of a 'correction' factor, $(1+C^2/n)^{-3/2}$,
which can be very large.
The square of the coefficient of variation defines a scale for $n$: when 
$n \gg C^2$, the law of large numbers approximately holds, whereas when 
$n \ll C^2$, the law of large numbers grossly overestimates 
$\Snt$. If the initial lognormal distributions is broader, $C^2$ is larger
 and, thus, larger $n$'s are required for the law of large numbers 
to apply.
We analyze now more precisely the small $n$ and large $n$ behaviours.

For $n=1$, \eq{e3.4.9} gives $S_1^\mathrm{t} \simeq e^{\mu-\sigma^2}$ 
which is, as it should be, the exact expression for the typical 
value $x^\mathrm{t}$ of a single lognormal term (see \eq{e2.5}).
For small $n$, we obtain 
\begin{equation}
n\ll C^2: \quad \Snt \simeq n^{5/2} \frac{\average}{C^3},
\label{e3.4.10}
\end{equation}
\ie a much faster dependence on $n$ then in the usual law of large numbers;
this evokes a L\'evy flight with exponent $\alpha=2/5$ (see \eq{e3.1.9}).
For large $n$, the expression \eq{e3.4.9} expands into
\footnote{The subleading term $\left(3 C^2 \average / 2\right)$ may 
not be the best approximation for $n \gg C^2$ (see \cite{BKL2002}).}:
\begin{equation}
n\gg C^2: \quad \Snt \simeq n \average - \frac{3}{2} C^2 \average.
\label{e3.4.11}
\end{equation}

The practical consequences of these expressions appear clearly on the 
sample mean $Y_n$:
\begin{equation}
Y_n \defin \frac{S_n}{n}.
\label{e3.4.11bis}
\end{equation}
\Eq{e3.4.10} and \eq{e3.4.11} give the typical sample mean $\Ynt$:
\begin{subequations}
\begin{equation}
n\ll C^2: \quad \Ynt \simeq \left( \frac{n}{C^2} \right)^{3/2} \average 
\end{equation}
\begin{equation}
n\gg C^2: \quad \Ynt \simeq \average - \frac{3}{2} \frac{C^2}{n} \average.
\end{equation}
\end{subequations}
Thus, for small systems $\left(n\ll C^2\right)$, one has $\Ynt \ll \average$.
In other words, {\it the sample mean of a small system does not typically yield the average 
value}. For instance, if $\sigma^2 = 4$, $Y_1^\mathrm{t} \simeq \average /400$.
This is important, \eg for tunnel junctions \cite{DRB2002} and contradicts 
common implicit assumptions \cite{Cho1963, Hur1966}. For large
systems ($n\gg C^2$), one recovers the average value. However, the correction 
to the average value decreases slowly with $n$, as $1/n$, and might be 
measurable even for a relatively large $n$%
\footnote{ This may explain why anomalous scaling effects have been observed 
in tunnel junctions as large as $10 \times 10 \mu\mathrm{m}^2$ \cite{Kel1999, 
LWP2002}.}.
Thus, macroscopic measurements may give access to microscopic fluctuations,
which is important for physics applications. Usually, microscopic fluctuations 
average out so that they
can not easily be extracted from macroscopic measurements. This property,
often taken for granted, comes from the fast convergence of sums $S_n$
to the law of large numbers asymptotics, which only occurs with narrow 
distributions.

\begin{figure}
\includegraphics[scale=0.33,angle=-90]{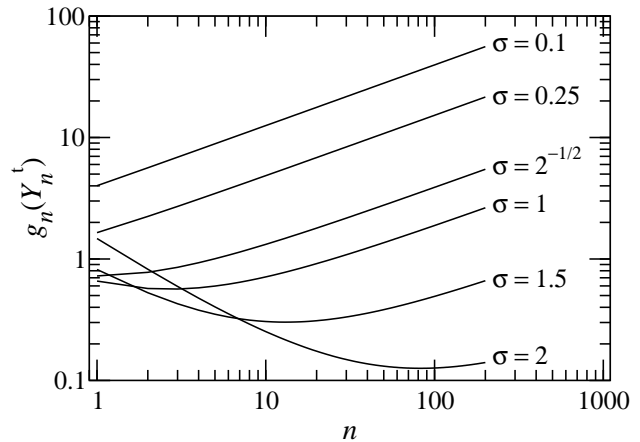}
\caption{Peak height $\gnstar(\Ynt)$ of the distribution of the sample mean 
 $Y_n=S_n/n$. Initial lognormal distribution: $\LN(\mu=0,
\sigma^2)$. 
For narrow lognormal distributions 
($\sigma^2 < 1/2$), $\gnstar(\Ynt)$ always increases with $n$ (normal 
behaviour). For broader lognormal distributions ($\sigma^2 > 1/2$), 
$\gnstar(\Ynt)$ presents an unusual decrease with $n$ at small $n$
indicating that the peak of $\gnstar(Y_n)$ broadens, even if its far
tail becomes lighter as usual.}
\label{fig3_4_1}
\end{figure}

We consider now the peak height $\gnstar(\Ynt)$ of the distribution
\begin{equation}
\gnstar(Y_n) \defin n \fnstar(S_n)
\label{e3.4.12bis}
\end{equation}
of the sample mean $Y_n$ (\fig{fig3_4_1}). 
Combining \eq{e2.9} and \eq{e3.4.1} via \eq{e3.4.5} and \eq{e3.4.7} gives:
\begin{equation}
\gnstar(\Ynt) = \frac{1+C^2/n}{\sqrt{2\pi} \average \sqrt{\ln (1+C^2/n)}}.
\label{e3.4.12}
\end{equation}
A simple study, for the non trivial case $\sigma^2 > 1/2$, reveals that 
$\gnstar(\Ynt)$ {\it decreases}
from $n=1$ to $n=C^2/(e^{1/2}-1)$ $(>1)$ and then 
increases for larger values of $n$.
This echoes the non-monotonous dependence on $\sigma^2$ of the
peak height of a lognormal distribution $f$ (see \eq{e2.9} and related 
comments).
The increase at large $n$ simply corresponds to the narrowing of the 
distribution of $Y_n = S_n/n$ when $n$ increases, as predicted by the law
of large numbers. Moreover, the large $n$ expansion of \eq{e3.4.12} gives
$\gnstar(\Ynt) \simeq \frac{\sqrt{n}}{\sqrt{2\pi\var{x}}}$, which is the 
prediction of the central limit theorem, as it should be. On the other hand,
the decrease of $\gnstar(\Ynt)$ at small $n$ is less usual. The peak 
of $\gnstar(Y_n)$ is actually {\it broader} than the one of the 
unconvoluted distribution $g_1 = f$. This behaviour can be understood
in the following way. If the lognormal distribution $f(x)$ is  
\New{broad}
enough ($C^2\gg 1$), it presents at the same time a high and narrow peak
at small $x$ {\it and} a long tail at large $x$. The effect of 
convoluting $f$ with itself is first ($n<C^2/(e^{1/2}-1)$) to `contaminate'
the peak with the (heavy) tail. This results in a broadening and decrease
of the $\fnstar(S_n)$ peak which is strong enough to entail
a decrease of the $\gnstar(Y_n) = n \fnstar(S_n)$ peak.
On the contrary, when enough convolutions have taken place ($n>C^2/(e^{1/2}-1)$),
the shape parameter $\sigma_n^2$ (\eq{e3.4.5}) becomes small and the tail of 
$\fnstar$ becomes light. Under these circumstances, further convolution
mainly `mixes' the peak with itself. This results in a broadening and
decrease of the $\fnstar(S_n)$ peak which is weak enough to allow an increase 
of the $\gnstar(Y_n) = n \fnstar(S_n)$ peak.

The small $n$ decrease of $\gnstar(Y_n)$ has physical consequences. There is a 
range of sample sizes, corresponding to $n<C^2/(e^{1/2}-1)$ for which 
the precise determination of the typical values becomes {\it more difficult}
when the sample size increases. This is a striking effect of the broad character
of the lognormal distribution%
\footnote{
This effect is also obtained for other broad distributions like, for
example, the L\'evy stable law $L_\alpha(x)$ with index $0< \alpha<1$ such that
$\average = \infty$. From L\'evy's generalized central limit theorem, 
the distribution of $S_n/n^{1/\alpha}$ is $L_\alpha$ itself
so that the distribution $l^{n*}(S_n/n)$ is
$n^{1-1/\alpha} L_\alpha(n^{1-1/\alpha} S_n/n)$.
As $\alpha<1$, one has $1-1/\alpha <0$ and the peak height of 
$l^{n*}(S_n/n)$ decreases with $n$.
}%
.
On the contrary, for narrow distributions, the determination of the 
typical value becomes more accurate as the sample size increases.

At last, we examine the compatibility of the obtained $\fnstar$ with the 
central limit theorem by studying the distribution $\hnstar(Z_n) = \fnstar
(S_n) \diff S_n /\diff Z_n$ of the usual rescaled random variable $Z_n$:
\begin{equation}
Z_n \defin \frac{S_n - n\average}{\sqrt{n~ \var{x}}}.
\label{e3.4.13}
\end{equation}
Simple derivations using \eq{e2.8} and \eq{e2.7m} lead to
\begin{eqnarray}
\hnstar(Z_n) \simeq \frac{C}{\sqrt{2\pi n \ln \left( 1 + \frac{C^2}{n} 
	\right) } \left( 1 + \frac{CZ_n}{\sqrt{n}} \right)}
	\nonumber \\
	\exp \left\lbrace \frac{-\left[ \ln \left( 1 + \frac{CZ_n}{\sqrt{n}} \right) 
			+ \frac{1}{2} \ln \left( 1 + \frac{C^2}{n} \right) 
		\right]^2}{2 \ln \left( 1 + \frac{C^2}{n} \right)} \right\rbrace.
\end{eqnarray}
For $n\gg C^2$ and $|C Z_n/\sqrt{n}| \ll 1$, one has 
$\ln (1+C^2/n) \simeq C^2/n$ and $\ln (1 + C Z_n/\sqrt{n}) \simeq 
C Z_n/\sqrt{n} - (C Z_n)^2/2n$, which gives:
\begin{equation}
\hnstar(Z_n) \simeq \frac{1}{\sqrt{2\pi}\left( 1 + \frac{CZ_n}{\sqrt{n}} \right)} \;
   \exp \left\lbrace \frac{-\left[ Z_n + \frac{C}{2\sqrt{n}} \left(1-Z_n^2\right) 
			\right]^2}{2} \right\rbrace .
\label{e3.4.15}
\end{equation}
Clearly, the central limit theorem is recovered%
\footnote{ The convergence to the central limit theorem can also be
derived less formally if one requires only the leading order of $\hnstar(Z_n)$. 
Indeed, \eq{e3.4.5} implies $\sigma_n^2 \simeq C^2/n$
when $n\to \infty$. Thus, $\sigma_n^2 \to 0$ when $n \to \infty$, 
\eq{ea1.6} applies: when $n\to \infty$, 
$\fnstar \simeq \LN (\mu_n, \sigma_n^2) \simeq 
\mathrm{N} \left(e^{\mu_n}, (\sigma_n e^{\mu_n})^2 \right) 
\simeq \mathrm{N} (n\average, n\var{x})$ since $\mu_n \simeq \ln(n\average)$
(see \eq{e3.4.7}). This agrees with the central limit asymptotics of 
\eq{e3.4.16}.
}:
\begin{equation}
\hnstar(Z_n) \to \frac{1}{\sqrt{2 \pi}} e^{-Z_n^2/2} \quad \mathrm{when} \quad 
	n \to \infty,
\label{e3.4.16}
\end{equation}
consistently with the strategy defined in \Section{s3.2} (see \eq{e3.2.2}).
Moreover, the square of the coefficient of variation appears in \eq{e3.4.15}
as the convergence scale of $\fnstar$ to the central limit theorem. As shown
in \eq{e3.4.9}, $C^2$ is also the convergence scale of $\Snt$ 
to the law of large numbers.

\subsection{Case of very broad lognormal distributions}
\label{s3.5}
We consider here the case $\sigma^2 \gg 1$ of very broad 
lognormal distributions. To treat this complex case, we will proceed through
different steps, in a more heuristic way than in the previous cases. 

The first step is to assume that the sums $S_n$ are typically dominated by the 
largest term $M_n$, if $n$ is not too large (see \eq{e3.1.4} and \Section{s3})
\footnote{Estimating the typical sum $S_n$ is then, in principle, an extreme
value problem; however, usual extreme value theories \cite{EKM1997} apply only
for irrelevantly large $n$ such that $S_n \simeq M_n$ is no longer valid.}.
Thus, the distribution function of $S_n$, defined as the probability that 
$S_n < x$ and denoted as $\Proba{S_n}{x}$, is approximately equal to the 
distribution function of $M_n$, denoted as $\Proba{M_n}{x}$:
\begin{equation}
\sigma^2 \gg 1 : \Proba{S_n}{x} \simeq \Proba{M_n}{x}.
\label{e3.5.1}
\end{equation}
As $M_n$ is the largest term of all $x_i$'s, $M_n < x$ is equivalent to 
$x_i < x$ for all $i = 1, \ldots, n$. Thus,
\begin{equation}
\Proba{M_n}{x} = \Proba{x_1}{x} \times \cdots \times 
\Proba{x_n}{x} = \left[ \Repart \right]^n
\label{e3.5.2}
\end{equation}
where $\Repart \defin \int_0^x{f(x') \diff x'}$ is the distribution function 
of the initial lognormal distribution. This implies
\begin{equation}
\Proba{S_n}{x} \simeq \left[ \Repart \right]^n
\label{e3.5.3}
\end{equation}
\footnote{A similar expression can be found without justification in 
\cite{ScY1982}, eq.~(16). A numerical study of this expression is presented in 
\cite{BAM1995}.}. By definition, the typical sum $\Vnt{S}$ is given by 
$\diff^2{\Proba{S_n}{x}} / \diff{x^2} = 0$, which, from \eq{e3.5.3}, leads
to:
\begin{equation}
-\left( \sigma + y_n \right) \sqrt{2\pi} \Phi\left(y_n\right) +
\left(n -1\right) e^{-y_n^2/2} = 0
\label{e3.5.4}
\end{equation}
where $y_n \defin \left( \ln \Vnt{S} - \mu \right) / \sigma$ and 
$\Phi\left(y\right) \defin \left( 2\pi\right)^{-1/2} \int_{-\infty}^y{e^{
-u^2/2} \diff u}$ is the distribution function of the standard normal
distribution $\N\left(0, 1 \right)$. This equation has no exact explicit 
solution. However, as $y_1 = -\sigma \ll -1$ (use \eq{e2.5} with 
$S_1^{\mathrm t} = \xt$), let us assume that $y_n \ll -1$ also for $n > 1$. 
Then we can approximate $\Phi\left(y_n\right)$ by $\Phi\left(y_n\right) \simeq 
-e^{-y_n^2/2} / \sqrt{2\pi} y_n$ (see, \eg \cite{AbS1972}, chap. 26). This 
leads to a linear equation on $y_n$, giving $y_n \simeq -\sigma/n$, valid for
$y_n \ll -1$, \ie $n < \sigma$. Finally, one has:
\begin{equation}
\sigma^2 \gg 1, n < \sigma : \Vnt{S} \simeq e^{\mu - \sigma^2/n}.
\label{e3.5.5}
\end{equation}
For $n = 1$ this expression is exact. When $n$ increases till 
$n = \sigma^2$, \eq{e3.5.5} gives an unusually fast, exponential 
dependence on $n$ that is in contrast with, \eg the $n^{5/2}$ 
dependence obtained for $\sigma^2 \lesssim 1$ and $n \ll C^2$ (\eq{e3.4.10}).
Unfortunately,
when $n$ becomes larger, \eq{e3.5.5} is qualitatively wrong. Indeed, it implies 
$\Vnt{S}/n \to e^{\mu}/n \to 0$ instead of $\Vnt{S}/n \to \average$ as 
predicted by the law of large numbers.

The second step, improving \eq{e3.5.5}, consists in combining \eq{e3.5.5} with
a cumulant constraint. We assume that $\fnstar \simeq \LN\left( 
\mu_n, \sigma_n^2 \right)$ as in \Section{s3.4} for all $n = 2^j$, 
with $j = 1, 2, \ldots$ and that the typical sum $S_{2^{j+1}}^{\mathrm t}$ is 
$e^{\mu_{2^j} - \sigma_{2^j}^2/2}$ as in \eq{e3.5.5} since $S_{2^j}$ is 
considered as lognormal. We use these assumptions and 
$\langle S_{2^{j+1}} \rangle = 2^{j+1} \average$ to determine induction 
relations between $\left( \mu_{2^{j+1}}, \sigma_{2^{j+1}}^2\right)$ and 
$\left( \mu_{2^j}, \sigma_{2^j}^2\right)$, which leads to:
\begin{subequations}
\label{e3.5.6}
\begin{equation}
\sigma_{2^j}^2 = \left( \frac{2}{3}\right)^j \sigma^2 + 2 
\left[ 1 - \left( \frac{2}{3}\right) \right] \ln 2,
\label{e3.5.6a}
\end{equation}
\begin{equation}
\mu_{2^j} = \mu + \left( \frac{\sigma^2}{2} - \ln2\right) 
\left[ 1 - \left( \frac{2}{3} \right)^j \right] + j \ln 2.
\label{e3.5.6b}
\end{equation}
\end{subequations}
The typical sum is then
\begin{equation}
\Vnt{S} \simeq n \average \exp\left[- \frac{3}{2} \frac{\sigma^2}{n^{\ln 
\left( 3/2\right)/\ln 2}} - 3 \ln 2 \left(1-\frac{1}{n^{\ln 
\left( 3/2\right)/\ln 2}} \right) \right].
\label{e3.5.7}
\end{equation}
\Eq{e3.5.7} is still exact for $n = 1$ and it clearly improves on \eq{e3.5.5} 
for large $n$. Indeed, when $n \to \infty$, $\Vnt{S}/n$ no longer tends to 0. 
However, $\Vnt{S}/n$ tends to $\average / 8$ instead of $\average$, 
which is the signature of a leftover 
problem. This comes from the assumptions that 
$\fnstar \simeq \LN\left( \mu_n, \sigma_n^2 \right)$, which may be correct for large 
$n$ (small $\sigma_n^2$) but is excessive for small $n$ (large $\sigma_n^2$), 
and that $S_{2^{j+1}}^{\mathrm t} \simeq e^{\mu_{2^j} - \sigma_{2^j}^2/2}$, 
which is correct for small $n = 2^j$ (large $\sigma_n^2$) but is excessive for 
large $n$ (small $\sigma_n$).

The third step, in order to cure the main problem of \eq{e3.5.7}, is to wildly 
get rid of the last term in the exponential which prevents $\Vnt{S}$ from 
converging to $n \average$ at large $n$, which does not affect the validity 
for $n = 1$:
\begin{equation}
\sigma^2 \gg 1: \quad \Vnt{S} \simeq n \average \exp\left[- \frac{3}{2} \frac{\sigma^2}{n^{\ln 
\left( 3/2\right)/\ln 2}}  \right].
\label{e3.5.8}
\end{equation}
\New{We have tried to empirically improve this formula by looking for a 
better exponent $\alpha$ than $\ln \left( 3/2\right)/\ln 2$ for 
$\sigma^2 \in \left[ 0.25, 16 \right]$. Unfortunately, no single $\alpha$ value 
is adequate for all $\sigma$'s. \Eq{e3.5.8} with $\alpha = 
\ln \left( 3/2\right)/\ln 2$ stands up as a good compromise for the 
investigated $\sigma$-range.}

\section{Range of validity of formulae}
\label{s4.1}
In this section we proceed to the numerical determination of the range of 
validity of the three theoretical formulae given by \eq{e3.3.7}, \eq{e3.4.9} 
and \eq{e3.5.8} for the typical sum of $n$ lognormal terms.

\begin{figure}
\includegraphics[scale=0.33,angle=-90]{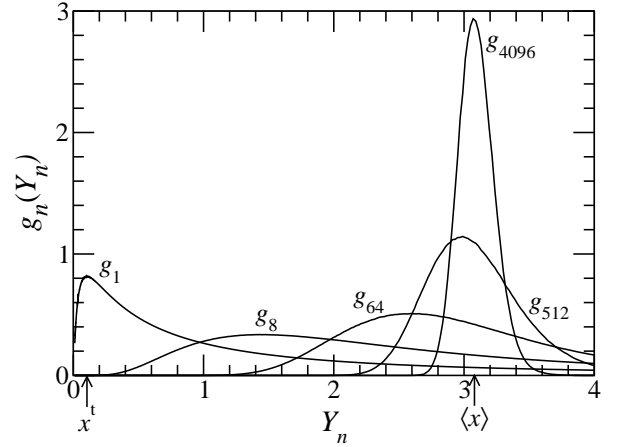}
\caption{Distributions $\gnstar(Y_n)$  of the sample mean $Y_n$ 
for an initial lognormal with $\mu =0$ and $\sigma = 1.5$.}
\label{fig4.1.1}
\end{figure}

\begin{figure}
\includegraphics[scale=0.33,angle=-90]{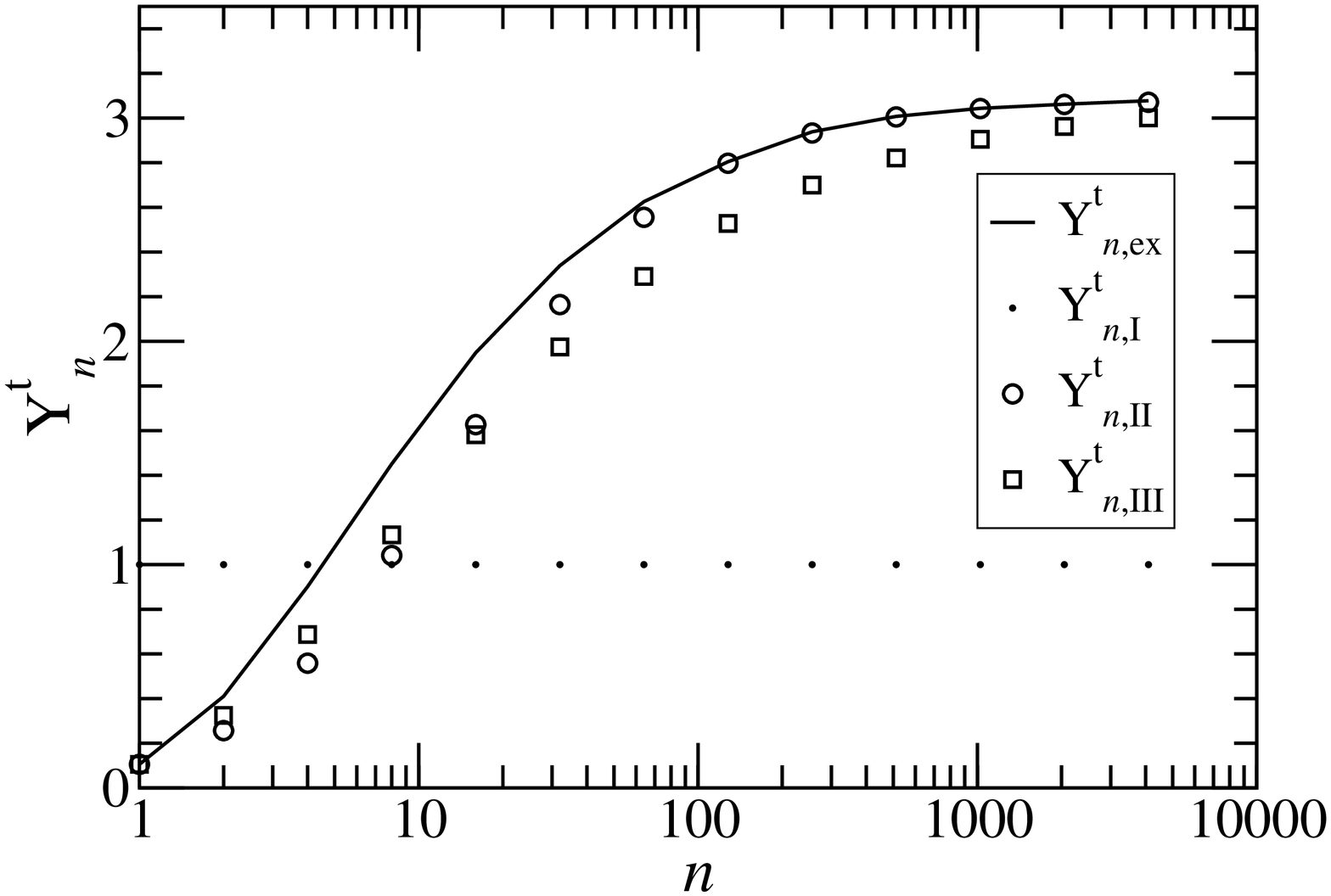}
\caption{$\Numerical{Y}{n}{t}$ (solid line) for an initial 
$\sigma = 1.5\ \left(\mu = 0\right)$ as well as $\YnI$ ($\sigma^2 \ll 1$, dots), $\YnII$ 
($\sigma^2 \lesssim 1$, circles) and $\YnIII$ ($\sigma^2 \gg 1$, squares).}
\label{fig4.1.2}
\end{figure}

In order to fulfil this task, the typical sample mean 
$\Vnt{Y}$ (\eq{e3.4.11bis}) instead of $\Vnt{S}$ will be used. This 
has the advantage of showing only the discrepancies to the mean 
value without the obvious proportionality of $\Vnt{S}$ on $n$ resulting from 
the law of large numbers. The values of  $\Vnt{Y}$ computed using the three
theoretical formulae \eq{e3.3.7}, \eq{e3.4.9} and \eq{e3.5.8} are called 
$\YnI$, $\YnII$  and $\YnIII$ respectively:
\begin{subequations}
\label{e4.1.0}
\begin{equation}
\sigma^2 \ll 1: \quad  \YnI = e^\mu,
\label{e4.1.0.a}
\end{equation}
\begin{equation}
\sigma^2 \lesssim 1: \quad \YnII = \average \left( 1 + \frac{C^2}{n} \right)^{-3/2},
\label{e4.1.0.b}
\end{equation}
\begin{equation}
\sigma^2 \gg 1: \quad \YnIII = \average \exp \left[ -\frac{3}{2} \, 
\frac{\sigma^2}{n^{\ln(3/2)/\ln2}} \right].
\label{e4.1.0.c}
\end{equation}
\end{subequations}
The exact typical sample mean,
derived from Monte Carlo generation \footnote{Standard numerical 
integration techniques to estimate the $n$-fold 
convolution of $f$ are impractical for broad distributions. On the contrary, 
the Monte Carlo scheme can naturally handle the coexistence of small and large 
numbers \cite{Bar1997}.}
of the distributions $\gnstar(Y_n)$, is called $\Numerical{Y}{n}{t}$. 
 Enough Monte-Carlo draws ensure 
negligible statistical uncertainty.
As an example, we show in \fig{fig4.1.1} the obtained distributions 
$\gnstar(Y_n)$ for $\mu = 0$ and $\sigma = 1.5$.
Notice that $\Numerical{Y}{n}{t}$ moves from $\Numerical{Y}{1}{t} = \xt =
e^{-\sigma^2} \simeq 0.11$ 
to $\Numerical{Y}{\infty}{t} = \average = e^{\sigma^2/2} \simeq 3.08$.
To determine the $\Numerical{Y}{n}{t}$'s, shown as solid line in 
\fig{fig4.1.2}, the absolute maximum of $\gnstar(Y_n)$ is obtained by 
parabolic least square fits performed on the $\log/\log$ representation of 
each distribution\footnote{A lognormal 
distribution reduces to a parabola in its $\log/\log$ representation.}.
Moreover, in the latter figure, we also 
show $\YnI$ (dots), $\YnII$ (circles) and $\YnIII$ (squares).

To determine the validity range of the theoretical formulae, we define 
two error estimators. The first one is the {\it maximum relative error} 
$\delta_{\mathrm {rel, (I,\ II,\ or\ III)}}$, \ie~ the maximum deviation 
referred to the minimum between $Y_{n, \mathrm{(I,\ II,\ or\ III)}}^\mathrm{t}$ 
and $\Numerical{Y}{n}{t}$, which is defined as follows: 
\begin{equation}
\deltareli \defin  \max\left( 
\frac{ \Yni - \Numerical{Y}{n}{t} } { \Numerical{Y}{n}{t} },
  \frac{ \Numerical{Y}{n}{t} - \Yni }{\Yni}; n = 1, 2, \ldots 
\right)
\label{e4.1.1}
\end{equation}
and can be transformed into:
\begin{equation}
\deltareli = \max\left[ e^{\left|\ln\left( \frac{\Yni} 
{\Numerical{Y}{n}{t}} \right) \right|} - 1; n = 1, 2, \ldots \right].
\label{e4.1.1bis}
\end{equation}
The second one is the {\it maximum scale error} 
$\delta_{\mathrm {scale, (I, II\ or\ III)}}$, \ie~the maximum
deviation in magnitude referred to the total amplitude of the phenomenon:
\begin{equation}
\delta_{\mathrm {scale}, i} \defin \max\left[ \left| \frac{ 
\ln\left(\Yni / \Numerical{Y}{n}{t} \right)}
{ \ln\left(\Numerical{Y}{\infty}{t} / \Numerical{Y}{1}{t} \right)} \right|; 
n = 1, 2, \ldots \right].
\label{e4.1.2}
\end{equation}
Using \eq{e2.5} for $\Numerical{Y}{1}{t}$ and \eq{e3.1.2} for 
$\Numerical{Y}{\infty}{t}$, $\delta_{\mathrm {scale}, i}$ boils down to:
\begin{equation}
\delta_{\mathrm {scale}, i} = \max\left[ \left| \frac{2 \ln\left(
\Yni / \Numerical{Y}{n}{t} \right)}
{ 3 \sigma^2} \right|; n = 1, 2, \ldots \right].
\label{e4.1.2bis}
\end{equation}
Remark that $\delta_{\mathrm {rel}, i} = \exp \left( \frac{3 \sigma^2}{2}
 \delta_{\mathrm {scale}, i} \right) - 1$. 
The first step for computing $\delta_{\mathrm {rel}, i}$ and 
$\delta_{\mathrm {scale}, i}$ is thus to find the value of $n$ for which 
$\left| \ln \left( \Yni / \Numerical{Y}{n}{t} \right) 
\right|$ is maximum. For the data shown in \fig{fig4.1.2}, we find 
$n = 1$ for \eq{e4.1.0.a}, $n = 4$ for \eq{e4.1.0.b} and $n = 4$ for 
\eq{e4.1.0.c}, which gives $\delta_{\mathrm {rel, I}} = 849\%
$ 
($\delta_{\mathrm {scale, I}} = 67\%
$), 
$\delta_{\mathrm {rel, II}} = 61\%
$ ($\delta_{\mathrm {scale, II}} = 14\%
$)
and $\delta_{\mathrm {rel, III}} = 31\%
$ 
($\delta_{\mathrm {scale, III}} = 8\%
$).

\begin{figure}
\includegraphics[scale=0.33,angle=-90]{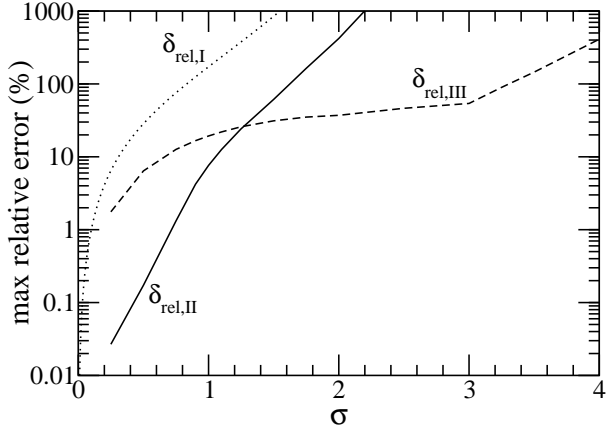}
\caption{Maximum relative errors $\delta_{\mathrm {rel}, i}$ as functions of 
$\sigma$.}
\label{fig4.1.3}
\end{figure}

\begin{figure}
\includegraphics[scale=0.33,angle=-90]{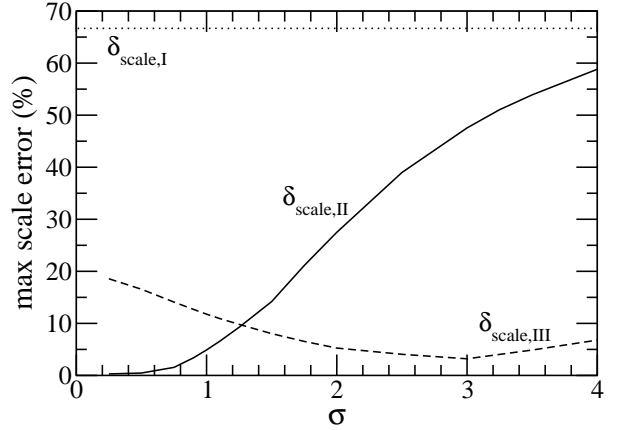}
\caption{Maximum scale errors $\delta_{\mathrm {scale}, i}$ as function of 
$\sigma$.}
\label{fig4.1.4}
\end{figure}

To work out the dependences of $\delta_{\mathrm {rel}, i}$
(\fig{fig4.1.3}) and $\delta_{\mathrm {scale}, i}$ (\fig{fig4.1.4}) as
functions of $\sigma$,  the same kind of calculation is 
performed for $\sigma \in (0,4]$ which is the relevant range for the chosen
physics applications.
The dotted lines representing $\delta_{\mathrm{rel, I}}$ and 
$\delta_{\mathrm {scale, I}}$ show that the first theoretical formula
is the least accurate in the explored $\sigma$ range. However, for its domain
of application, $\sigma^2 \ll 1$, the error is acceptable for 
$\delta_{\mathrm{rel, I}}$ ($\delta_{\mathrm{rel, I}} \simeq \sigma^2$,
see \footnote{
The quantity $\delta_{\mathrm{rel, I}}$ can be computed analytically. 
Indeed, $\YnI = e^\mu$ does not depend on $n$ and $\Numerical{Y}{n}{t}$ 
is bound by $\Numerical{Y}{1}{t} = e^{\mu-\sigma^2}$ and 
$\Numerical{Y}{\infty}{t} = e^{\mu+\sigma^2/2}$. Thus 
$\delta_{\mathrm{rel, I}} = \max ( e^{\sigma^2} - 1, e^{\sigma^2/2} - 1) = e^{\sigma^2} - 1$. 
This implies $\delta_{\mathrm{scale, I}} = 2/3$
 for any $\sigma^2$, in agreement with \fig{fig4.1.4}.
}
).
Indeed, $\delta_{\mathrm{rel, I}} \lesssim 7\%
$ for $\sigma \in [0, 0.25]$
which, in turn, means that lognormal distributions are quasi-Gaussian in 
this range (see shaded area in \fig{fig3.3}). 
The solid lines representing $\delta_{\mathrm{rel, II}}$ and 
$\delta_{\mathrm{scale, II}}$ show that the second theoretical formula is 
the most accurate in the range $0\leq \sigma \lesssim 1.25$ giving 
$\delta_{\mathrm{rel, II}}\lesssim 30 \%
$ and
$\delta_{\mathrm{scale, II}}\lesssim 10 \%
$. Note that good tunnel junctions
fall within this $\sigma$ range. The broken lines representing 
$\delta_{\mathrm{rel, III}}$ and $\delta_{\mathrm{scale, III}}$ show that 
the third theoretical formula is the most accurate for $\sigma \gtrsim 1.25$
and is reasonably accurate for $\sigma \lesssim 1.25$. Note that,
for $\sigma = 4$, the maximum relative error 
$\delta_{\mathrm{rel, III}} \simeq 400 \%
$ appears quite high. However,
when the error is referred to the total amplitude of the scaling,
as given by $\delta_{\mathrm{scale, III}}$, it is only $7\%
$.

Importantly, the observed ranges of validity of the three different formulae 
are consistent with the strategies of approximation used to derive these 
formulae. This provides an a posteriori confimation of the theoretical
analysis presented in the paper.

\section{A striking effect: scaling of the sample mean and of its inverse}
\label{s4.2}
\New{In general, if a function is increasing, its inverse is decreasing. 
What happens if one considers the typical values of a random variable and of 
its inverse ? Does one have :
\begin{equation}
z_n^t \nearrow \iff \left( 1/z_n \right)^t \searrow \quad ?
\label{e4.2.0}
\end{equation}
While this is intuitively true for narrow distributions, it may fail 
for broad distributions.}

This problem arises in electronics, where it is customary to study the product 
$R \times A$ of the 
device resistance $R$ by the device size $A$. One usually checks that 
$R \times A$ does not depend on $A$, otherwise this dependence is taken 
as the indication of edge effects.
The resistance $R$ being the inverse of the 
conductance can be represented by $1 / S_n$ where $S_n$ is the sum of $n$ 
independent conductances. The size $A$ of the 
system is proportional to $n$. Hence, one has:
\begin{equation}
R \times A \propto \frac{n}{S_n} = \frac{1}{Y_n},
\label{e4.2.1}
\end{equation}
where $Y_n$ is the sample mean of conductances. We have shown that the typical 
 \New{$Y_n$} increases with the sample size 
(see eqs.~(\ref{e4.1.0})), if 
conductances are lognormally distributed. Hence, \New{$R \times A$ being 
proportional to the inverse of $Y_n$,} {\it one naively expects a 
decrease of the typical value of $R \times A$ with \New{$n \propto A$}}.

\New{What do the results presented in this paper imply for the {\it typical 
value of $R \times A$}?}
Let us do the correct calculation in the case $\sigma^2 \lesssim 1$, relevant 
for good tunnel junctions. As $\fnstar\left( S_n \right) \simeq \LN \left( 
\mu_n, \sigma_n^2 \right)\left( S_n \right)$, the distribution of $1 / Y_n$ 
is:
\begin{equation}
\LN \left( 
-\mu_n + \ln n, \sigma_n^2 \right)\left( 1 / Y_n \right)
\label{e4.2.2}
\end{equation}
(see \Section{s2}). 
The typical sample mean inverse is thus, using eqs.~(\ref{e3.4.5}) 
and~(\ref{e3.4.7}):
\begin{equation}
\sigma^2 \lesssim 1: \quad \left(1 / Y_n \right)^{ \mathrm t} 
\simeq \frac{1}{\average \left( 1 + \frac{C^2}{n} \right)^{1/2}} .
\label{e4.2.3}
\end{equation}
Thus just as $\Vnt{Y}$, $\left(1 / Y_n\right)^{ \mathrm t}$ increases 
with the sample size!

This counterintuitive result epitomizes the paradoxical behaviour of some 
broad distributions. Moreover, this can be a possible explanation for the 
anomalous scaling of $R \times A$ observed for small magnetic tunnel junctions 
\cite{LWP2002}.

\section{Conclusion}
\label{s5}
We have studied the typical sums of $n$ lognormal random variables. Approximate 
formulae have been obtained for three different regimes of the shape parameter 
$\sigma^2$. Table~\ref{table1} summarizes these results with their ranges of 
applicability.
\begin{table}
\caption{Range of applicability of the different formulae. 
\New{Errors are measured by $\delta_{\mathrm{rel}}$ and $\delta_{\mathrm{scale}}$,} 
see \Section{s4.1} 
for details.}
\label{table1}       
\begin{tabular}{lccc}
\hline\noalign{\smallskip}
$\Vnt{S}$ & $\sigma$ range & $\delta_{\mathrm{rel}}$ & $\delta_{\mathrm{scale}}$ \\
\hline\noalign{\smallskip}
$n e^{\mu}$ & $\left[ 0, 0.25 \right]$ & $\le 7\%
$ & $\le 67\%
$ \\
$n \average \frac{1}{\left( 1 + \frac{C^2}{n} \right)^{3/2}}$ & $\left[ 0, 1.25 \right]$ & $\le 30\%
$ & $\le 10\%
$ \\
$n \average \exp\left[- \frac{3}{2} \frac{\sigma^2}{n^{\ln \left( 3/2 \right)/\ln 2}}  \right]$ & $\left[ 1.25, 4 \right]$ & $\le 400\%
$ & $\le 7\%
$ \\
\noalign{\smallskip}\hline
\end{tabular}
\end{table}
These results are relevant up to $\sigma \lesssim 4$; for larger $\sigma$, one 
may apply the theorems in \cite{BBG2002} and \cite{BKL2002}.

The anomalous behaviour of the typical sums has been related to the 
broadness of lognormal distributions. For large enough shape parameter 
$\sigma^2$, the behaviour of lognormal sums is non trivial. It reveals 
properties of broad distributions at small sample sizes and properties of 
narrow distributions at large sample sizes with a {\it slow transition} between the 
two regimes. Counter-intuitive effects have been pointed out like the decrease 
of the peak height of the sample mean distribution with the sample size and 
the fact that the typical sample mean and its inverse do not vary with the 
sample size in opposite ways.
Finally, we have shown that the statistical effects arising 
from the broadness of lognormal distributions have observable consequences for 
moderate size physical systems.

\acknowledgments
We thank O.E. Barndorff-Nielsen, G. Ben Arous, J.-P. Bouchaud, A. Bovier,  
H. Bulou \New{and F. Ladieu} for discussions. F. B. thanks A. Crowe, K. Snowdon and the 
University of Newcastle, where part of the work was done, for their hospitality.

\appendix
\section{Approximation of {\it narrow} lognormal distributions by 
normal distributions and vice versa}
\label{a1}
As seen in Section~\ref{s2}, the lognormal probability distribution 
$f(x) = \LN(\mu, \sigma^2)(x)$ is mostly concentrated in the interval 
$\left[e^{\mu}e^{-\sigma}, e^{\mu}e^{\sigma}\right]$. If $\sigma \ll 1$, 
this range is small and can be rewritten as:
\begin{equation}
e^{\mu} \left( 1 - \sigma \right ) \lesssim x 
\lesssim e^{\mu} \left( 1 + \sigma \right ).
\label{ea1.1}
\end{equation}
Thus it makes sense to expand $f(x)$ around its typical value $e^{\mu}$ by
introducing a new random variable $\epsilon$ defined by:
\begin{equation}
x \defin e^{\mu} \left( 1 + \epsilon \right),
\label{ea1.2}
\end{equation}
where $\epsilon$ is a
random variable on the order of $\sigma$:
\begin{equation}
-\sigma \lesssim \epsilon \lesssim \sigma.
\label{ea1.3}
\end{equation}
As $\sigma \ll 1$, this entails $\left| \epsilon \right| \ll 1$.
Expanding the lognormal distribution $f(x)$ of \eq{e1.1} in powers of
$\epsilon$ leads to:
\begin{eqnarray}
f(x) & \simeq & \frac{1}{\sqrt{2 \pi \sigma^2} e^{\mu}} \left( 1 - \epsilon + \epsilon^2
+ \cdots \right) \nonumber \\
 & & \exp \left( - \frac{\epsilon^2}{2 \sigma^2} +
\frac{\epsilon^3}{2 \sigma^2} + \cdots \right)
\label{ea1.4}
\end{eqnarray}
The dominant term gives $f(x) \simeq \frac{1}{\sqrt{2 \pi \sigma^2} e^{\mu}} \exp
\left( - \frac{\epsilon^2}{2 \sigma^2} \right)$, thus using \eq{ea1.2}:
\begin{equation}
f(x) \simeq \frac{1}{\sqrt{2 \pi \left( \sigma e^{\mu} \right)^2}} 
\exp \left[ -\frac{ \left( x -
e^{\mu} \right)^2}{2 \left( \sigma e^{\mu} \right)^2} \right]
\label{ea1.5}
\end{equation}
In other words, a {\it narrow} lognormal distribution is well approximated by a
normal distribution:
\begin{equation}
\sigma \ll 1 : \LN\left( \mu, \sigma^2 \right) \simeq N\left( e^{\mu}, \left(
\sigma e^{\mu} \right)^2 \right)
\label{ea1.6}
\end{equation}

More intuitively, the Gaussian approximation of narrow lognormal
distributions $\LN\left( \mu, \sigma^2 \right)\left( x \right)$ can be 
inferred from the underlying Gaussian random variable $y$ with 
distribution $N\left( 0, 1 \right)\left( y \right)$, with 
$x = e^{\mu + \sigma y}$. Since $\left| y \right|
\simeq 1$ and $\sigma \ll 1$, one has $\left| \sigma y \right| \ll 1$ and,
 thus, $x \simeq e^{\mu} \left( 1 + \sigma y \right)$. Consequently, $x$ 
being a linear transformation of a Gaussian random variable, is itself 
normally distributed according to 
$N\left( e^{\mu}, \left( \sigma e^{\mu} \right)^2 \right)$, in
agreement with \eq{ea1.6}. 

Conversely, a {\it narrow} $\left( \sigma \ll \mu \right)$ 
Gaussian distribution $N\left( \mu, \sigma^2\right)$ can be approximated by 
a lognormal distribution:
\begin{equation}
\sigma \ll 1 : N\left( \mu, \sigma^2 \right) \simeq \LN\left( \ln \mu, \left(
\sigma / \mu \right)^2 \right).
\label{ea1.7}
\end{equation}

For completeness, one can
easily show that {\it any} Gaussian distribution 
$N\left( \mu, \sigma^2 \right)$
can be approximated by a three parameter lognormal
distribution 
$\LN\left( \ln\left( \mu + A \right), \left( \frac{\sigma}{\mu + A}
\right)^2, A \right)$ where $A$ is any number such that $A + \mu \gg \sigma$.
The probability density of the three parameter lognormal distribution is 
 $\LN\left(\mu, \sigma^{2}, A \right) = \frac{1}{\sqrt{2 \pi
\sigma^{2}} \left( x - A \right)} \exp\left\lbrace - \frac{\left[ \ln\left( x -
A \right) - \mu \right]^2}{2 \sigma^{2}} \right\rbrace$ for $x > A$ and $0$
otherwise.

\bibliographystyle{apsrev}
\bibliography{bibFB}

\newpage

\end{document}